\documentclass[prb,onecolumn, superscriptaddress]{revtex4}
\usepackage[pdftex]{graphicx}
 \usepackage{amsmath}
\usepackage[T1]{fontenc}
\usepackage{amsfonts}
\usepackage{lipsum}
\usepackage{amssymb, mathrsfs}
\usepackage{braket}
\usepackage{subfigure}
\def\beq{\begin{equation}}
\def\eeq{\end{equation}}
\def\bsp{\begin{split}}
\def\esp{\end{split}}
\def\bea{\begin{eqnarray}}
\def\eea{\end{eqnarray}}
\def\ba{\begin{array}}
\def\ea{\end{array}}

\def\dg{\dagger}

\def\lb{\left(}
\def\rb{\right)}

\def\l.{\left.}
\def\r.{\right.}

\def\ra{\rangle}
\def\la{\langle}

\def\bo{\bold{k}}

\begin{document}

\title{Dirac Magnon Nodal Loops in Quasi-2D Quantum Magnets}
\author{S. A. Owerre}
\email{sowerre@perimeterinstitute.ca}

\affiliation{Perimeter Institute for Theoretical Physics, 31 Caroline St. N., Waterloo, Ontario N2L 2Y5, Canada.}

\begin{abstract}
 \textbf {In this report, we propose a new concept of  one-dimensional (1D) closed lines of Dirac magnon nodes in two-dimensional (2D) momentum space of quasi-2D quantum magnetic systems. They are  termed  ``2D Dirac magnon nodal-line loops''.   We utilize  the bilayer honeycomb ferromagnets with intralayer coupling ${\bf J}$ and interlayer coupling ${\bf J_L}$, which is realizable  in the honeycomb chromium  compounds CrX$_3$ (X $\equiv$ Br, Cl, and I). However, our results can also exist in other layered quasi-2D quantum magnetic systems. Here, we show that the magnon bands of the bilayer honeycomb ferromagnets overlap  for ${\bf J_L\neq 0}$  and form 1D closed lines of Dirac magnon nodes in 2D momentum space. The 2D Dirac magnon nodal-line loops are topologically protected by inversion and time-reversal symmetry.  Furthermore, we show that they are robust against weak Dzyaloshinskii-Moriya   interaction ${\bf \Delta_{DM}< J_L}$ and possess chiral magnon edge modes.}
\end{abstract}
\maketitle

The experimental observations of topological insulators \cite{top3,top4} and topological semimetals \cite{Xu,lv1,lv2} in electronic systems  have inspired a great interest in condensed matter physics. Consequently, this has led to the re-examination   of band structures in bosonic systems. One of the areas of recent interest is the topological magnon bands in insulating ordered quantum magnets with inversion symmetry breaking  \cite{lifa,alex4,alex5a,shin1,sol,sol1,kkim},  which allows a Dzyaloshinskii-Moriya (DM) spin-orbit  interaction (SOI) \cite{dm,dm2}.  The study of topological magnetic spin excitations in quasi-two-dimensional (2D) quantum magnetic systems  is currently an active research field in condensed matter physics both theoretically and experimentally. Topological magnonic systems  are expected to open the next frontier of condensed matter, because they are potential candidates towards magnon spintronics and magnon thermal devices \cite{magn}.   Unlike electronic charge  particles,  magnons are charge-neutral bosonic quasiparticles and they do not experience a Lorentz force and do not have conduction and valence bands. However,  a temperature gradient can induce a heat current and the  Berry curvature induced by the DM SOI acts as an effective magnetic field in momentum space. This leads to a thermal version of the anomalous Hall effect characterized by a temperature dependent thermal Hall conductivity   as predicted theoretically \cite{alex0,alex5,alex2, alex22} and subsequently observed experimentally \cite{alex1a,alex1,alex6}.

   Recently, Mook   et al.~\cite{mok1} have  shown that the magnon bands in the three-dimensional (3D) anisotropic pyrochlore ferromagnets without DM SOI  form a magnon analogue  of 3D electronic nodal-line semimetals \cite{aab2,aab3,aab4,aab5}. However,  the concept of the quasi-2D magnon nodal-line has never been proposed  in quantum magnetic systems. In principle, there are more quantum magnetic materials in  quasi-2D forms than 3D forms. In addition, quasi-2D quantum magnetic systems are simpler  to study and manipulate  both theoretically and experimentally. Therefore, it is interesting to seek for a 2D analogue of Dirac magnon  nodal-lines.  Nevertheless, the concept of 2D electronic Dirac nodal-line semimetals has remained elusive for a while. It was recently put forward and it is predicted to exist in composite lattices such as  mixed  honeycomb and kagom\'e lattices \cite{lu}, mixed square lattice \cite{yang}, and 2D trilayers \cite{che}. In this case,  the mixed energy bands have the tendency to overlap in momentum space and form 1D closed lines of Dirac nodes in 2D momentum space. A magnonic analogue of these systems should be of interest in the study of quantum magnetism both theoretically and experimentally.

 In this report, we propose a 2D Dirac magnon nodal loop or 1D closed lines of Dirac magnon nodes in 2D momentum space  as a direct analogue of 2D electronic Dirac nodal-line semimetals \cite{yang,lu,che}.  We utilize  the  honeycomb bilayer ferromagnet as an example,  which can be realized in the hexagonal chromium  compounds CrX$_3$ (X $\equiv$ Br, Cl, and I) with honeycomb lattices coupled by small interlayer interaction \cite{dav0,dav,dav1,foot, dejon}. However, the general idea of 1D closed lines of Dirac magnon nodes in 2D momentum space can be easily extended to  other quasi-2D lattice structures with more than two sublattices in the unit cell. We show here that in the realistic regime with the interlayer  coupling ($J_L$) smaller than the intralayer coupling ($J$), {\it i.e.}  $J_L<J$,  there are Dirac magnon nodal loops centred at the corners of the Brillouin zone. We show that the 2D Dirac nodal-line loops are topologically protected by the  $\mathbb Z_2$ invariance of the parity eigenvalue resulting  from the presence of inversion symmetry.  In principle, the DM SOI with strength $\Delta_{DM}$ can be allowed on the honeycomb lattice because of inversion symmetry breaking between the bonds of second-nearest neighbour sites \cite{sol}. With the inclusion of DM SOI, we show that the quasi-2D  Dirac magnon nodal-line loops are robust  for $\Delta_{DM}<J_L$. This is in stark contrast to  3D magnon nodal-lines for which the DM SOI transforms the magnon nodal-lines into magnon Weyl points \cite{mok1}.


\textbf{Results}.

\textbf{Model}. We consider the ferromagnetic spin Hamiltonian of a bilayer honeycomb lattice, given by 
\begin{align}
\mathcal H&=  -J\sum_{\la i,j\ra;\tau }{\bf S}_{i,\tau}\cdot{\bf S}_{j,\tau}- J_L\sum_{i}{\bf S}_{i}^T\cdot{\bf S}_{i}^B - H\sum_{i,\tau}S_{i,\tau}^z.
\label{h}
\end{align}
The first term describes the intralayer nearest-neighbour (NN) interaction on the top ($\tau=T$) and bottom ($\tau=B$) layers respectively, and the second term is the interlayer interaction.    The last term is the Zeeman magnetic field in units of $g\mu_B$ applied out of  the bilayer plane (in the $z$-direction). The bilayer honeycomb lattice can be stacked in two forms: AB-stacked and AA-stacked. In the former the top layer is slightly shifted from the bottom layer, whereas they are placed right above each other in the latter. We will assume the latter as the former does not possess the Dirac magnon nodal-line loops proposed in this report. Also, in the latter  the Hamiltonian \eqref{h} is a good approximation of the honeycomb ferromagnetic chromium compounds CrX$_3$ (X $\equiv$ Br, Cl, and I) \cite{dav0,dav,dav1,foot, dejon}. The intralayer coupling is ferromagnetic $J>0$ and the small interlayer coupling can be ferromagnetic $J_L>0$ as in X $\equiv$ Br and I \cite{dav0,dav,dav1} or antiferromagnetic $J_L<0$ as in X $\equiv$ Cl \cite{foot}. We note that CrCl$_3$ and CrBr$_3$ have an estimated value of $|J_L|/J\sim 0.004$  and  $J_L/J\sim 0.06$ respectively  \cite{dav0,dav}. These values are very close to the magnetic susceptibility measurements \cite{dejon}. As we will show later, the Dirac magnon-nodal lines exist for any value of $J_L\neq 0$. In the presence of a strong magnetic field, the antiferromagnetic coupled  layers can be mapped to the ferromagnetic coupled layers at the saturation field $H_s=2J_LS$, where $S$ is the value of the spin. Therefore, we will consider only ferromagnetic coupled layers.  Due to small interlayer coupling $J_L<J$, it is reasonable to consider the quasi-2D system.

\begin{figure}
\includegraphics[width=4in]{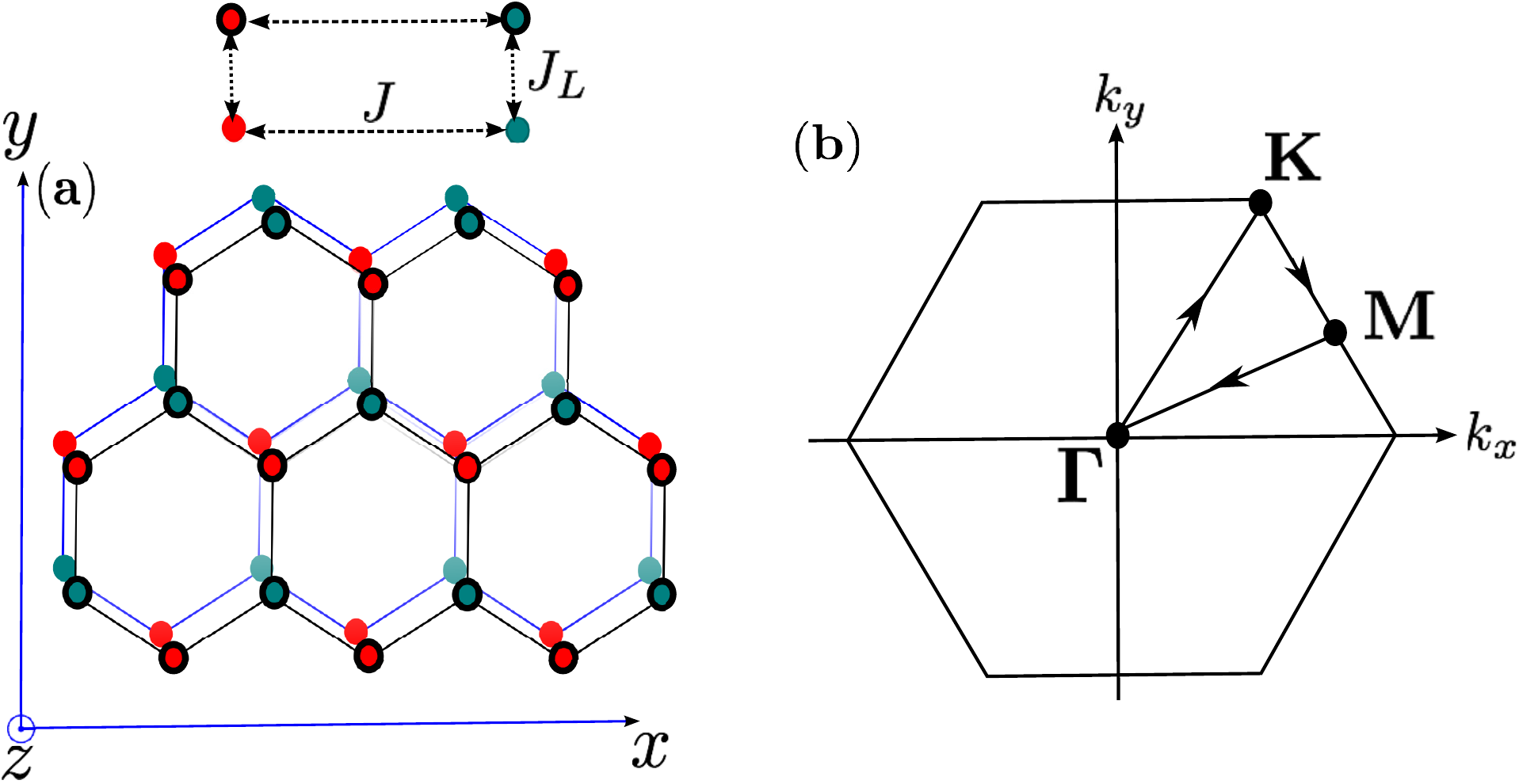}
\caption{\textbf{The honeycomb lattice and the Brillouin zone.} {\bf (a)} Schematic representation of AA-stacked bilayer honeycomb magnet on the plane perpendicular to the $z$-axis. {\bf (b)} Brillouin zone of the system with different paths. }
\label{pl1}
\end{figure}

\begin{figure}
\includegraphics[width=3.2in]{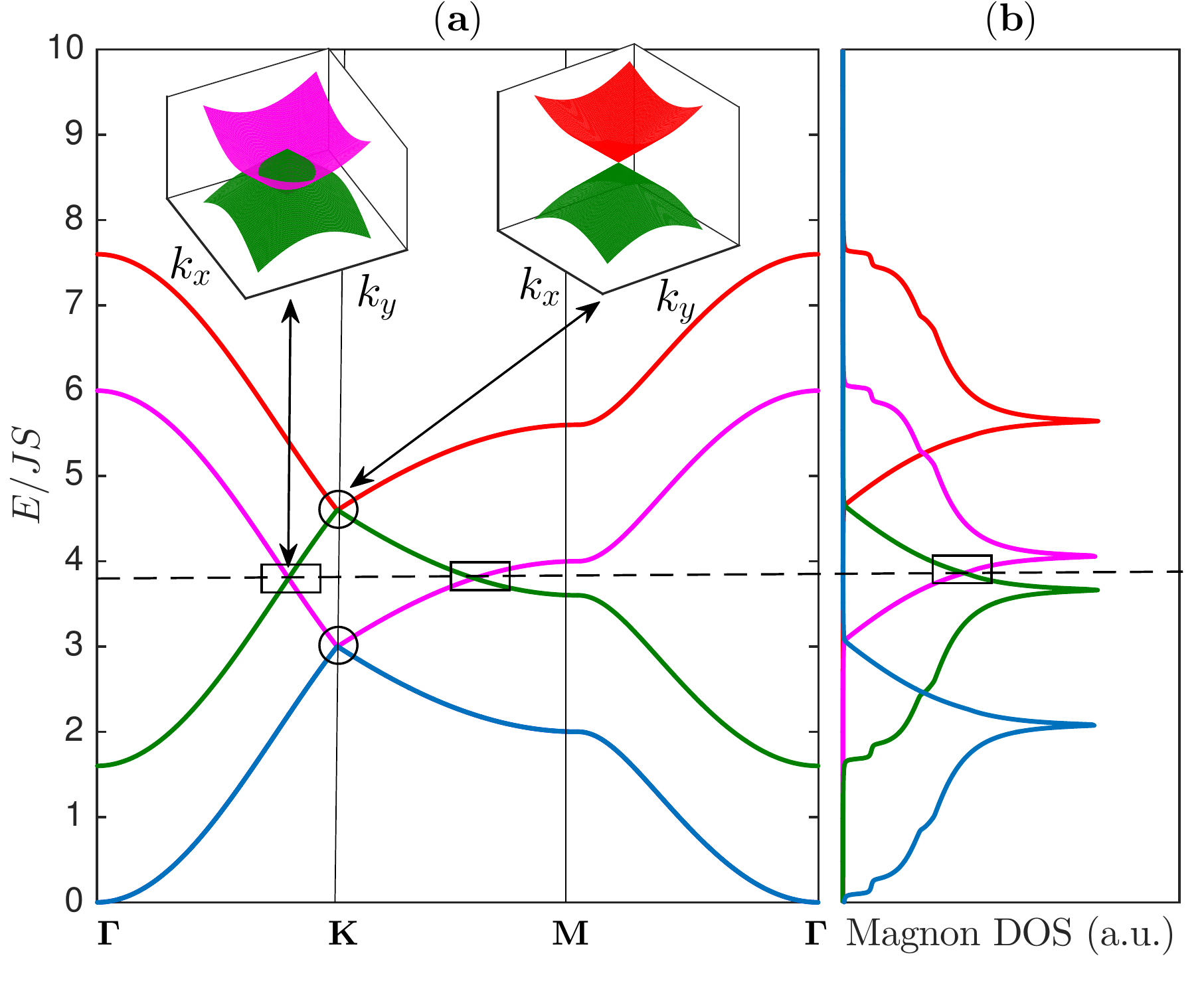}
\caption{\textbf{The magnon energy bands and magnon density of states (DOS).}  {\bf (a)} Magnon band structures with coexistence of magnon Dirac-nodal lines (rectangular boxes) and  magnon Dirac points (circles)  along the Brillouin zone paths in Fig.~\ref{pl1}{\bf (b)}. The dashed line corresponds to the energy of Dirac magnon-nodal lines at $E=E_0$.  The inset shows the 3D bands near  ${\bf K}$  for the Dirac magnon-nodal lines and  Dirac magnon points respectively as indicated by the same colour codes of the 2D bands. {\bf (b)} The corresponding magnon density of states (DOS). The plots are generated by setting  $J_L=0.8J$}
\label{pl2}
\end{figure}

\textbf{Magnon band structures}. 
 At low-temperatures the magnetic excitations of ferromagnetic materials can be measured by inelastic neutron scattering experiment \cite{dav1,alex6} and they correspond to  the quanta of spin waves also known as magnons.  The linearized Holstein-Primakoff (HP) \cite{hp} spin-boson mapping suffices in the low-temperature regime.   The corresponding magnon bands of Eq.~\eqref{h} are given by (see Methods)  
\begin{align}
E_\pm^\alpha(\bo)= E_0 \pm S[ |Jf_\bo| +(-1)^\alpha |J_L|],
\label{band}
\end{align}
where $E_0=3JS+ J_LS + H$ and $f_{\bo}= 1 + e^{-ik_a}+e^{-i(k_a +k_b)}$, $\hat a=\sqrt{3}\hat x$ and $\hat b= -\sqrt{3}\hat x/2 + 3\hat y/2$ with $k_a=\bo\cdot\hat a$ and $k_b=\bo\cdot\hat b$. Here,   the lattice constant is denoted by $a$ and $\alpha=1$ or $2$. The magnetic field $H$ only introduces a gap at the Goldstone mode at $\bo=0$ and therefore it has no effects on the Dirac magnon points or the Dirac magnon nodal-lines. Henceforth, we  set $H=0$.  The topmost band is $E_+^2$ and the bottommost band is $E_-^2$ and the two middle bands are $E_\pm^1$.    For theoretical purposes, we will vary the value of $J_L/J$ in order to magnify the Dirac magnon nodal-lines as well as the Dirac points. For instance, the magnon bands are depicted in Fig.~\ref{pl2}(a) along the Brillouin zone paths in Fig.~\ref{pl1}(b) for $J_L/J=0.8$.   The topmost band $E_+^2$ and the lower middle band $E_+^1$ form a Dirac magnon point at $\pm {\bf K}$. Similarly,  the lowest band $E_-^2$ and the upper middle band $E_-^1$ form a Dirac magnon point at $\pm {\bf K}$. This is very similar to single-layer honeycomb ferromagnet \cite{mag}. In other words, the Dirac magnon points do not necessarily require a bilayer form.
 
\begin{figure}
\includegraphics[width=3.2in]{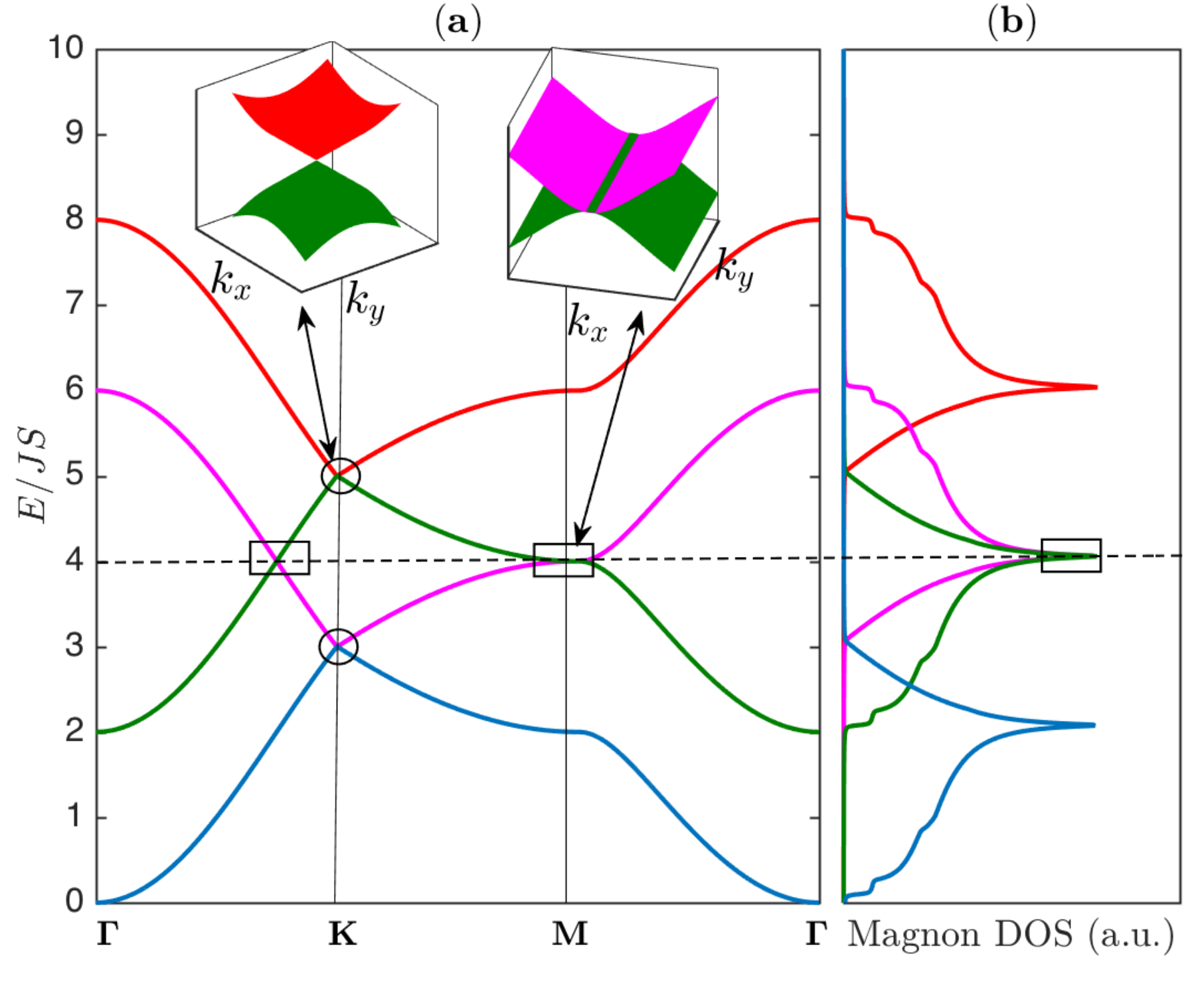}
\caption{\textbf{The magnon energy bands and magnon  density of states.}  {\bf (a)} Magnon band structures with coexistence of magnon Dirac-nodal lines (rectangular boxes) and  magnon Dirac points (circles)  along the Brillouin zone paths in Fig.~\ref{pl1}{\bf (b)}. The dashed line corresponds to the energy of Dirac magnon-nodal lines at $E=E_0$.  The inset shows the 3D bands near  ${\bf M}$  for the Dirac magnon-nodal lines and near ${\bf K}$ for the Dirac magnon points  respectively as indicated by the same colour codes of the 2D bands. {\bf (b)} The corresponding magnon density of states (DOS). The plots are generated by setting  $J_L=0.8J$.}
\label{pl3}
\end{figure} 

\begin{figure*}
\includegraphics[width=5in]{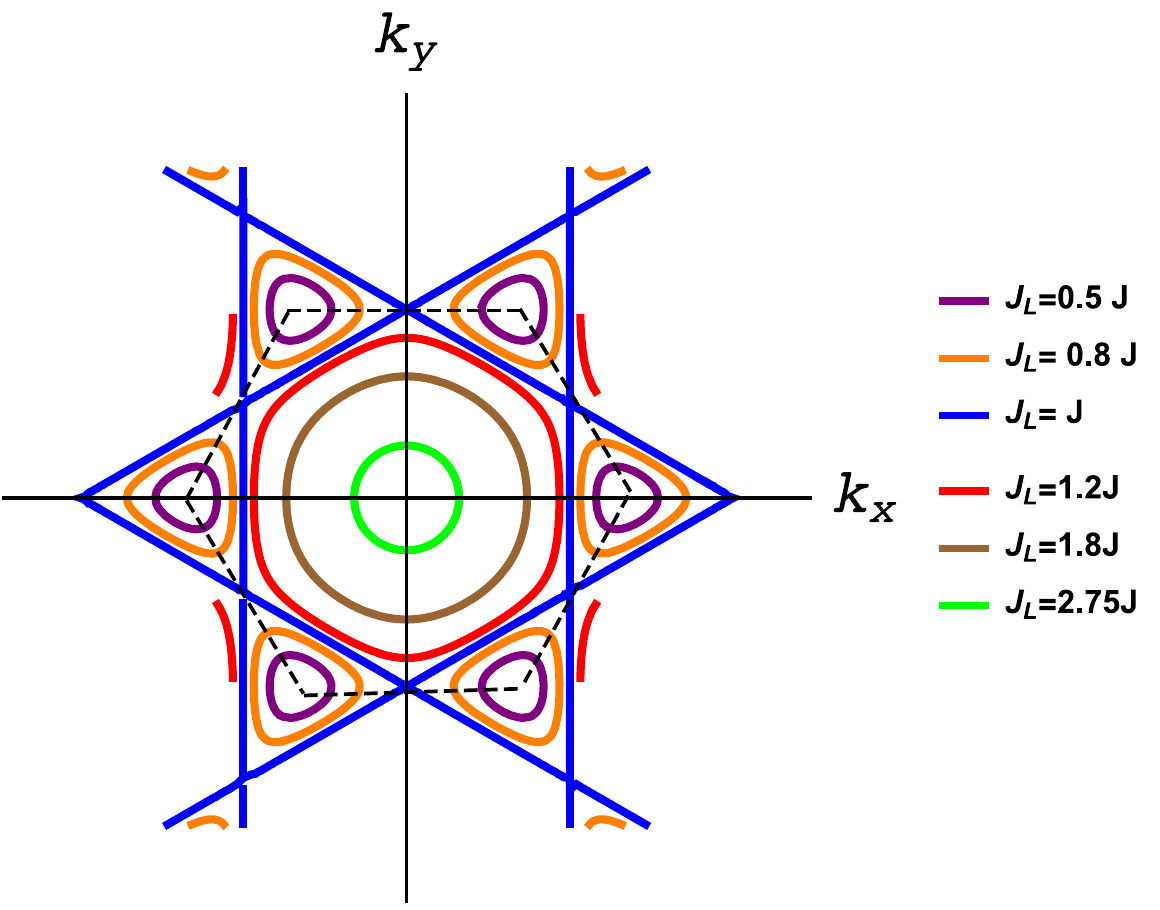}
\caption{Dirac magnon closed loops in momentum space for several values of  $J_L/ J$. The dashed black hexagon represents the Brillouin zone.}
\label{p4}
\end{figure*}

 \begin{figure}
\includegraphics[width=3.2in]{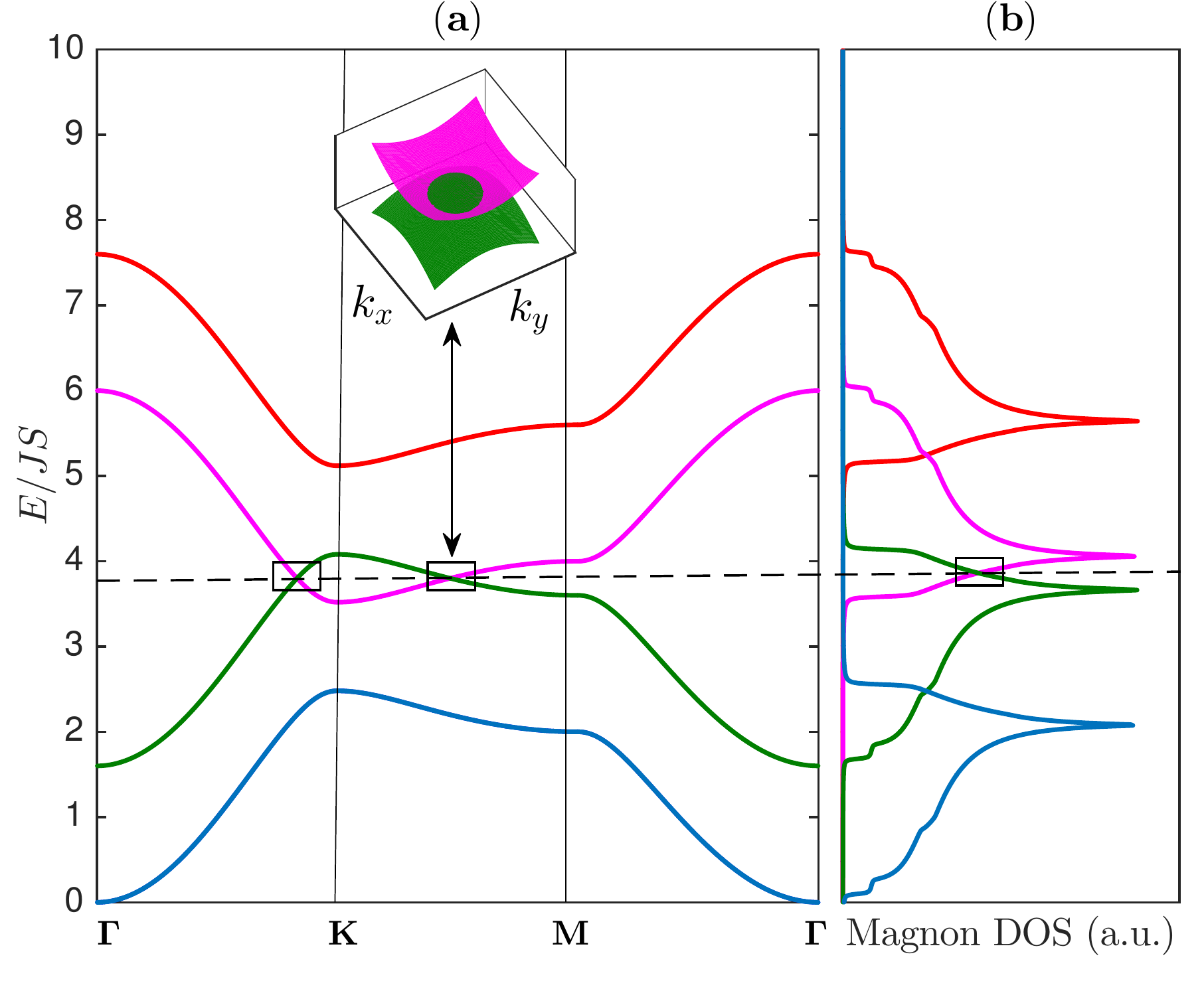}
\caption{The magnon energy bands and magnon density of states.  {\bf (a)} Magnon band structures with  Dirac magnon-nodal lines (rectangular boxes), but  gap Dirac magnon  points  along the Brillouin zone paths in Fig.~\ref{pl1}{\bf (b)}. The dashed line corresponds to the energy of Dirac magnon-nodal lines at $E=E_0$.  The inset shows the 3D band near ${\bf K}$ as indicated by the same colour codes of the 2D bands. {\bf (b)} The corresponding magnon density of states (DOS). The plots are generated by setting $\Delta_{DM}=0.1J, J_L=0.8J$.  }
\label{pl5}
\end{figure}

 \begin{figure}
\includegraphics[width=3.2in]{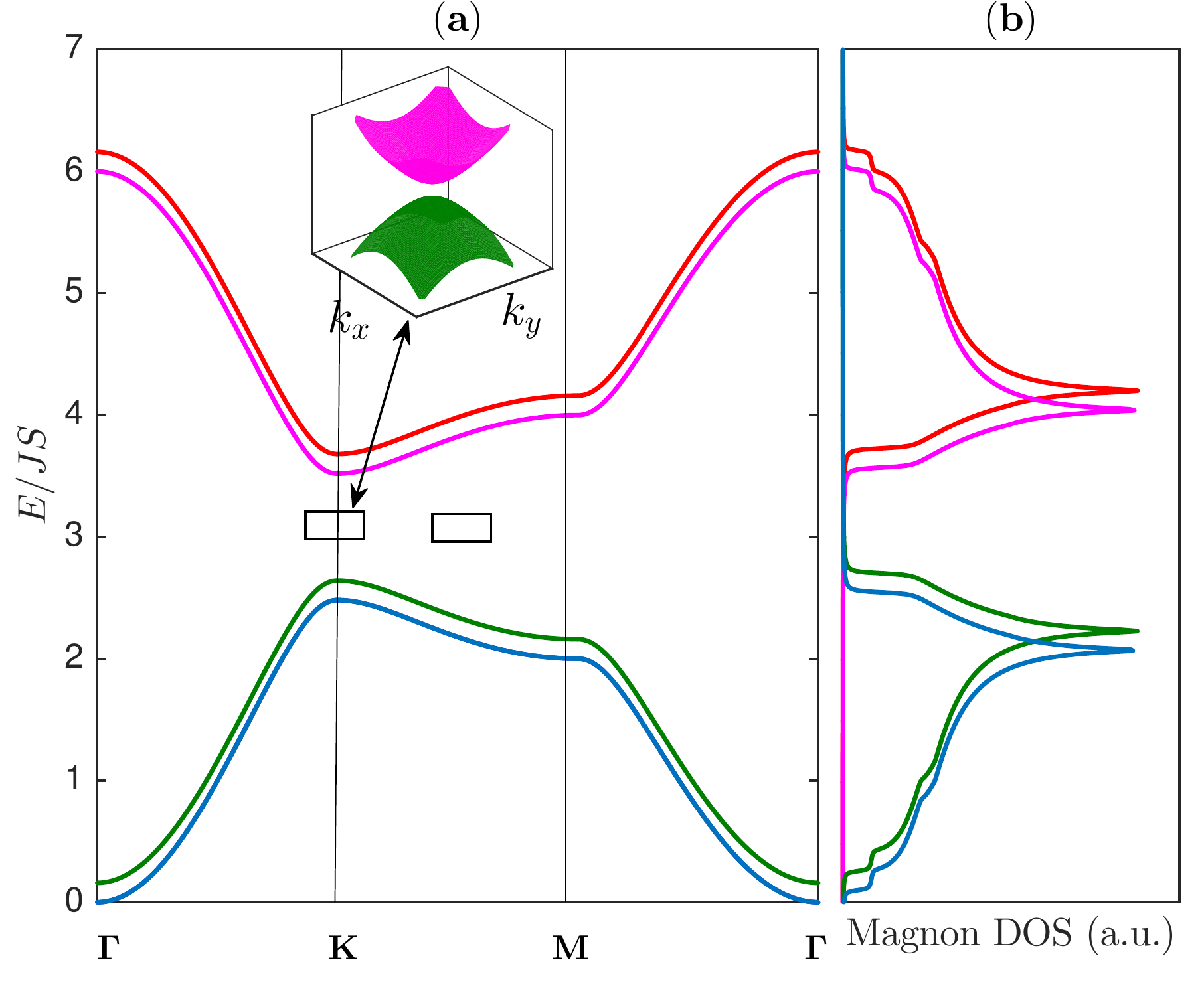}
\caption{The magnon energy bands and magnon  density of states.  {\bf (a)} Magnon band structures with gap Dirac magnon-nodal lines (rectangular box) as well as gap  Dirac magnon points along the Brillouin zone paths in Fig.~\ref{pl1}{\bf (b)}.  The inset shows the 3D band near ${\bf K}$ as indicated by the same colour codes of the 2D bands. {\bf (b)} The corresponding magnon density of states (DOS). The plots are generated by setting $\Delta_{DM}=0.1J, J_L=0.08J$.  }
\label{pl6}
\end{figure}

\begin{figure}
\includegraphics[width=3.2in]{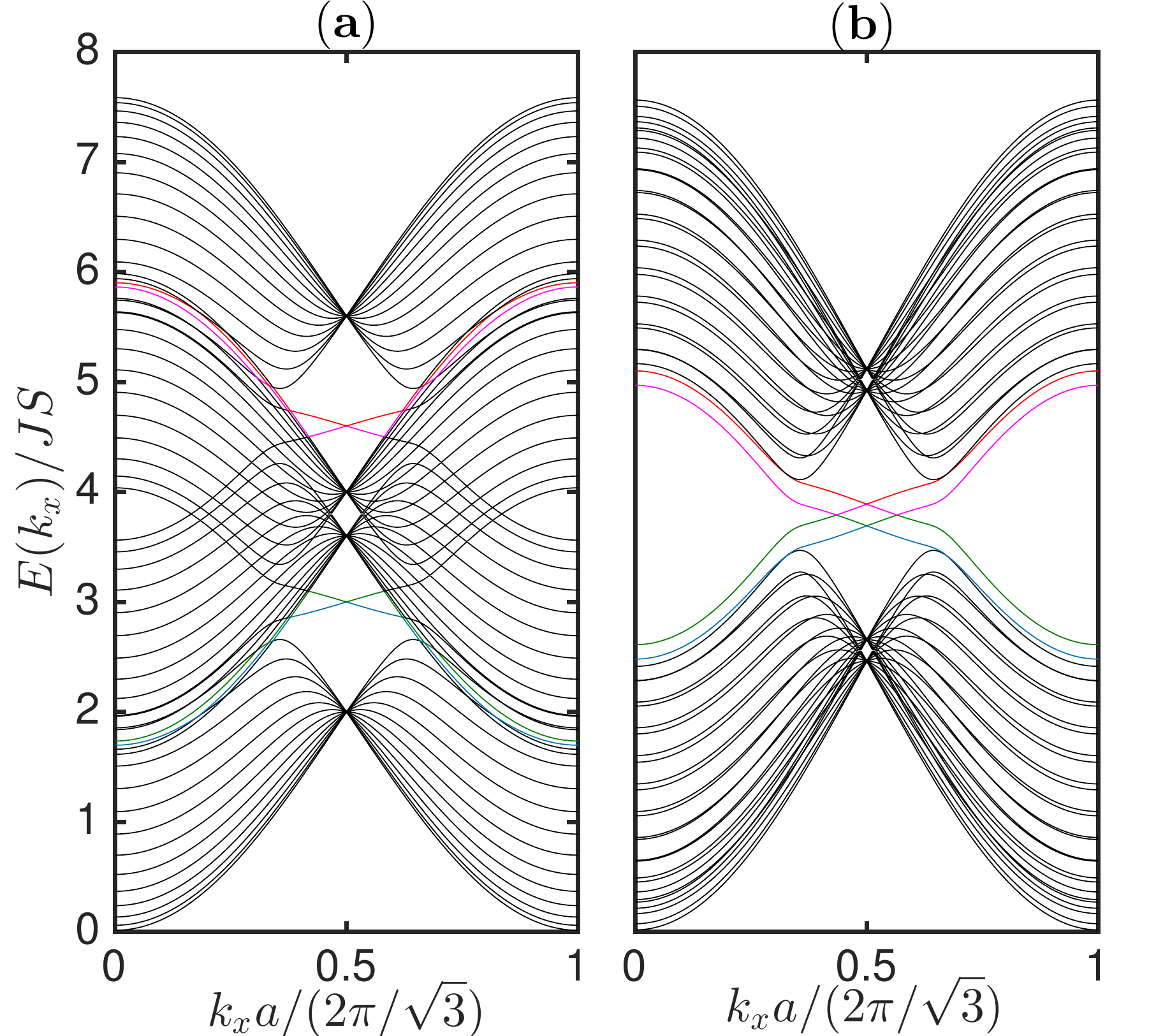}
\caption{Magnon bulk bands (black lines) and chiral magnon edge modes (coloured lines). A strip of width $N_y=14$  unit cells with open boundary conditions along $y$ and infinite along $x$ is used. The plots are generated by setting $\Delta_{DM}=0.1J$.  {\bf (a)} $J_L=0.8J$. {\bf (b)} $J_L=0.08J$.}
\label{edg}
\end{figure}

\textbf{Dirac magnon nodal-line loops}. The bilayer ferromagnets have another important features that are not possible in single-layer ferromagnetic form \cite{mag}. As evident from Fig.~\ref{pl2}(a) the middle magnon bands $E_\pm^1$ in the bilayer honeycomb ferromagnets overlap, say, along  ${\bf \Gamma}$--${\bf K}$ line in the Brillouin zone and they form  1D closed lines of Dirac nodes in 2D momentum space or 2D Dirac magnon nodal-line closed loops. They are different from the Dirac magnon points formed  by discrete band touching at the high symmetry $\pm {\bf K}$-points.  They are the exact analogues of 2D Dirac node-line semimetals in electronic systems that also exist in mixed lattices such as the honeycomb and kagom\'e lattices \cite{lu}, mixed square lattice \cite{yang}, and 2D trilayers \cite{che}. Indeed, for any value of $J_L \neq 0$ there exists 2D Dirac magnon nodal lines in the magnon band structures.  The magnon density of states (DOS) in Fig.~\ref{pl2}(b)  linearly cross between the bands forming the Dirac magnon nodal-loops and they are  distinguished from that of  Dirac magnon points.
As in electronic systems the 2D materials offer additional tunability which may not be available in 3D systems.  If the interlayer coupling  $J_L$ is considered as a tunable parameter we see from  Fig.~\ref{pl3}(a) that upon increasing  $J_L$ the Dirac magnon  points remain intact while the Dirac magnon nodal closed loops along ${\bf K}$--${\bf M}$ form a line node near the ${\bf M}$-point. The corresponding  magnon  density of states in Fig.~\ref{pl3}(b) also show a line of degeneracy consistent with the magnon bands.

One of the advantages of magnon bands in quasi-2D quantum magnets is that they offer an explicit analytical simplification, which may not be possible in 3D systems such as the pyrochlore ferromagnets \cite{mok1}. In this respect, the existence of Dirac magnon nodal-loops and Dirac magnon points are consequences of the magnon band structures and can be directly understood from Eq.~\ref{band}. In magnonic systems  both the Dirac magnon  points and the Dirac magnon nodal-loops  occur at finite energy $E_0$.  The solution of $E_\pm^\alpha=E_0$ determines the nodal-line degeneracy of the bands. There is no solution for the topmost and bottommost bands $E_\pm^2$, which is consistent with the absence of Dirac magnon closed loops. However, the zeros of the middle bands $E_\pm^1=E_0$ give the solution  where the two bands overlap in momentum space, {\it i.e.}, 2D Dirac magnon nodal-line closed loops. 

The main result of this report is depicted  in Fig.~\ref{p4}.  For $J_L<J$ the Dirac magnon nodal loops are centred at the corners of the Brillouin zone. At the  symmetric point $J_L=J$ there are additional  line nodes connecting the Dirac magnon nodal loops. Although $J_L>J$ is usually unrealistic, there are still Dirac magnon nodal loops in this regime as shown in Fig.~\ref{p4},  but now they are centred at the ${\bf \Gamma}$-point and shrink as $J_L$ increases and subsequently turn into a quadratic point node at the unrealistic limit critical value $J_L^c=3J$.

 To get an intuition about the development of Dirac magnon nodal loops we consider the  low-energy limit. Near ${\bf k}={\bf K}$ the zeros of $E_\pm^1=E_0$  gives
\begin{align}
q_x^2 +q_y^2 =  J_L^2/c_s^2,
\end{align}
where $c_s=3J/2$ and $\bold q=\bold{k}-{\bf K}$. The radius of the Dirac magnon nodal loops $|J_L/c_s|$  increases with $J_L$ consistent  with  Fig.~\ref{p4} for $J_L<J$. On the other hand, near ${\bf k}={\bf \Gamma}$ the zeros of $E_\pm^1=E_0$ gives
\begin{align}
q_x^2 +q_y^2 =  [J_L^2-(3J)^2]/c_0^2,
\end{align}
where $c_0=3J/\sqrt{2}$ and $\bold q=\bold{k}-{\bf \Gamma}$. As noted above  the Dirac magnon nodal loops shrink to a quadratic point node at the critical value $J_L=J_L^c$ as can be seen in  Fig.~\ref{p4}.

 By  linearizing $\mathcal H_\bo$ (see Methods) near the Dirac magnon points and the Dirac magnon nodal-line loops, and projecting onto the subspace associated with the middle bands, we obtain a 2D Dirac-like Hamiltonian
\begin{align}
\mathcal H(\bold q)= E_0 \bold{I}_{2\times 2}-v_0(\pm q_x\sigma_x +q_y\sigma_y),
\end{align}
where $v_0=c_s(c_0)$ is the spin wave velocity.
The linearized Hamiltonian has a Berry phase \cite{kai} defined as $
\gamma=\oint_{\mathcal C} \mathcal A(\bold q)\cdot d{ \bold q} $, where 
$\mathcal A(\bold q)$ is the Berry connection given by $\mathcal A(\bold q)=i\braket{\psi_{\bold q}|{\vec \nabla}_{\bold q}\psi_{\bold q}}$. The Berry phase is defined for a closed path encircling the Dirac point nodes in momentum space with $\gamma=\pi$, whereas $\gamma=0$ otherwise. Therefore the Dirac point nodes can be interpreted as topological defects in 2D momentum space. On the other hand, the 2D Dirac magnon nodal-line loops are topologically protected through the parity eigenvalues at the time-reversal-invariant momenta (TRIM) points due to the presence of inversion symmetry. In two dimensions there are four TRIM points at $\Gamma_i=(\boldsymbol{\Gamma}, 3\bold{M}'s)$ points, where the $\bold{M}$ points are the bisects of the Brillouin zone corners. The $\mathbb Z_2$ invariance $\nu$ is given by \cite{fu}
\begin{align}
(-1)^\nu=\prod_{i=1}^4\delta_i,
\end{align}
where $\delta_i=\prod_{n=1}^{2N}\xi_{n}(\Gamma_i)$, and $\xi_{n}(\Gamma_i)$ is the parity eigenvalue associated with the magnon bands forming the Dirac nodal-line loops. As shown in Methods $\delta_i=-\text{sgn}\lb d_1(\bo=\Gamma_i)\rb$, where $d_1(\bo)=-JS\text{Re} f_\bo$. We find that $\nu=1$ which yields a nonzero topological invariant $\mathbb Z_2=1$. This confirms the  topological protection of the odd number of Dirac nodal-line loops in the 2D Brillouin zone.


\textbf{Effects of  Dzyaloshinskii-Moriya interaction}. The honeycomb lattice can also allow a DM SOI due to  inversion symmetry breaking  at the midpoint of the bonds connecting the second-nearest neighbours which form a triangular plaquette.  The form of the DM SOI  on the honeycomb lattice  is given by \cite{sol} \bea \mathcal H_{DM}=\Delta_{DM}\sum_{\la \la i,j\ra\ra;\tau} \nu_{ij}{\hat z}\cdot{\bf S}_{i, \tau}\times{\bf S}_{j, \tau},\eea where $\nu_{ij}=\pm$ for hopping from right to left on each of the two-layer sublattices. The DM  SOI is imaginary in the spin-boson mapping and the resulting magnon band structures  are given by

\begin{align}
E_\pm^\alpha(\bo)= E_0 \pm S[ \sqrt{|Jf_\bo|^2 +m_\bo^2} +(-1)^\alpha J_L],
\label{band1}
\end{align}
where $m_\bo=-m_{-\bo}=2\Delta_{DM} [\sin k_a +\sin k_b-\sin(k_a+k_b)]$. The combination of ferromagnetic spontaneous magnetization and DM SOI breaks time-reversal symmetry macroscopically and the Dirac points are gapped  leading to a topological magnon insulator in single-layer honeycomb ferromagnets also known as the Haldane magnon insulator \cite{sol,kkim} with finite thermal Hall effect \cite{sol1}.

It is interesting to ask what happens to the Dirac magnon nodal loops and the Dirac magnon points as the DM SOI is taken into account in the bilayer system. We find that the Dirac magnon points are gapped due to broken  time-reversal symmetry. On the other hand,  the Dirac magnon nodal loops are not gapped by the DM  SOI in the realistic limit $\Delta_{DM}<J_L$ as shown in Fig.~\ref{pl5}; therefore they are robust in this regime. Nevertheless, the opposite limit $\Delta_{DM}>J_L$ has gap Dirac  magnon loops and Dirac  magnon points and the system becomes a full topological magnon insulator as depicted in Fig.~\ref{pl6}. These interesting features can be well-understood in the low-energy limit by expanding the magnon energy bands near ${\bf k}={\bf K}$. The zeros of $E_\pm^1=E_0$ give
\begin{align}
q_x^2 +q_y^2 = (J_L^2-\Delta_{DM}^2)/c_s^2,
\end{align}
where $\Delta_{DM}=3\sqrt{3}\tilde\Delta_{DM}$. Evidently, the presence of Dirac magnon nodal loop requires $\Delta_{DM} <J_L$. 

  Another interesting feature of this system is the existence of chiral magnon edge modes on the boundary. We have solved for the magnon edge modes using a strip geometry with open boundary conditions along the $y$-direction and infinite along $x$-direction.   As shown in Fig.~\ref{edg} there are four chiral magnon  edge modes in both regimes $\Delta_{DM}<J_L$ (a) and $\Delta_{DM}>J_L$ (b) with each band contributing a single edge mode. The topological protection of the magnon edge modes is encoded in the Chern numbers of the magnon bulk bands. For a gap  topological system the  Berry curvature is given by $\Omega_{\bo\pm}^\alpha= \lb\boldsymbol{\nabla}\times \boldsymbol{\mathcal A}_{\pm\bo}^\alpha\rb_z$, where  $\boldsymbol{\mathcal A}_{\pm\bo}^\alpha=i\braket{\psi_{\pm\bo}^\alpha|\boldsymbol{\nabla}|\psi_{\pm\bo}^\alpha}$ is the Berry connection  and $\psi_{\pm\bo}^\alpha$ are the eigenvectors of $\mathcal H_\bo$ given in Methods. The Berry phase $\gamma$ now defines a Chern number given by the integration of the Berry curvature over the Brillouin zone,
 \begin{equation}
\mathcal{C}_\pm^\alpha= \frac{1}{2\pi}\int_{{BZ}} d^2k~ \Omega_{\bo\pm}^\alpha.
\label{chenn}
\end{equation}
The Berry curvature has its maximum  at $\pm{\bf K}$. Therefore,  the major contributions to the  Chern numbers come from the states near the Dirac points. By direct numerical integration, we find that for the topmost and the upper middle bands $E_+^2$ and  $E_-^1$ respectively the Chern numbers are  $\mathcal C_\mp^{1(2)}=\pm 1$. Similarly, for the bottommost and the lower middle bands $E_-^2$ and  $E_+^1$ we find $\mathcal C_\mp^{2(1)}=\pm 1$. This is consistent with the number of chiral magnon edge modes in the system.

\textbf{Conclusion}. We have shown that the magnon bands in the quasi-2D quantum magnetic systems offer another new topological distinction --- the quasi-2D Dirac magnon nodal loop or 1D closed lines of Dirac magnon nodes in 2D momentum space. They are the direct analogues of 2D electronic Dirac nodal-line semimetals in  composite lattices  \cite{lu,yang}. The  Dirac magnon-nodal loops occur when two magnon bands overlap in momentum space as opposed to Dirac magnon points which occur when two magnon bands touch at discrete points at the high symmetry points of the Brillouin zone. We also showed that the quasi-2D Dirac magnon-nodal loops can exist even in the presence of DM SOI as opposed to 3D  counterparts on the pyrochlore ferromagnets \cite{mok1}.  It would be of interest to search for the existence of Dirac magnon-nodal loops in quantum magnetic systems via the  inelastic neutron scattering experiments.   Recently, many different topological  magnonic systems have been proposed; they include    Weyl magnon points  in 3D pyrochlore antiferromagnets  \cite{fei} and  ferromagnets \cite{ mok,su}, and  also Dirac points in quasi-2D  quantum magnetic systems  with DM SOI \cite{sol4,oku}, which  can be  categorized as the magnonic analogue of  2D Dirac semimetals \cite{ste}.  The present work has never been studied in any quasi-2D quantum magnetic system; hence it completes the magnonic analogues of electronic topological semimetals in 2D  systems. The present work is applicable to the honeycomb chromium compounds CrX$_3$ (X $\equiv$ Br, Cl, and I) \cite{dav0,dav,dav1,foot, dejon}. Therefore, the magnon bands of these honeycomb magnetic materials should be re-examined  experimentally in the context of topological magnonics. We note that the study of magnonic analogues of electronic topological semimetals is still developing. There are very few experiments \cite{alex1a,alex1,alex6} and lots of theoretical proposals have not been achieved experimentally. Because of the bulk sensitivity of inelastic neutron scattering methods, the chiral magnon edge modes have not been experimentally measured in any quantum magnetic system, but the bulk topological magnon bands have been realized in the kagom\'e lattice ferromagnet Cu(1-3, bdc) \cite{alex5a}. Therefore, edge sensitive methods should be applied in the study of topological magnonics. In this respect, it be would interesting to implement other experimental  techniques such as  light \cite{luuk} and electronic \cite{kha} scattering methods. We also note that the electronic properties of the honeycomb chromium compounds CrX$_3$ (X $\equiv$ Br, Cl, and I) have been studied by density functional theory and other methods \cite{sem1,sem2} and they possess semiconducting features.

\textbf{Methods} 
The formalism we implement in this report is the linear spin wave theory. This approximation is valid at low-temperature in the ordered regime when few magnons are thermally excited.  We assume that the spins are polarized along  the $z$-axis, which can be achieved by applying a magnetic field along this direction.  The linearized Holstein-Primakoff (HP) spin-boson mapping is given by

\begin{align} 
& S_{i,\tau}^{ z}= S-a_{i,\tau}^\dagger a_{i,\tau},
\\& S_{i,\tau}^{ y}=  i\sqrt{\frac{S}{2}}(a_{i,\tau}^\dagger -a_{i,\tau}),
\\& S_{i,\tau}^{ x}=  \sqrt{\frac{S}{2}}(a_{i,\tau}^\dagger +a_{i,\tau}),
\end{align} 
where $a_{i,\tau}^\dagger (a_{i,\tau})$ are the bosonic creation (annihilation) operators. There are four sites in the unit cell, {\it i.e.} $\tau=1,2$ on the top layer and $\tau=3,4$ on the bottom layer.  Next, we substitute the spin-boson transformation into Eq.~\ref{h} and Fourier transform the quadratic bosonic Hamiltonian which is similar to that of AA-stacked graphene \cite{dav2,dav3, davv}, apart from the constant energy shift $E_0=3JS+ J_LS+H$ that pushes the negative bands to positive bands as expected for  magnonic  (bosonic) systems. In other words, there is no conduction and valence bands in magnonic systems as the spin excitations are charge-neutral and obey Bose-Einstein statistics. Instead, the magnon bands are thermally populated at low temperatures.  
The momentum space Hamiltonian $\mathcal H_\bo$ can be written as
$\mathcal H =\sum_{\bold k}\psi^\dagger_{\bold k}\cdot \mathcal{H}_\bold k\cdot\psi_{\bold k},$
 where  $\psi_\bo^\dg=(a_{\bo,1}^\dg,a_{\bo,4}^\dg,a_{\bo,3}^\dg,a_{\bo,2}^\dg)$ is the eigenvectors. The momentum space Hamiltonian is given by
 \begin{align}
 \mathcal{H}_\bold k=E_0 I_\tau\otimes I_\sigma - J_L S \tau_x\otimes \sigma_x -JS\tau_x\otimes[\sigma_+ f_\bo + \sigma_- f_\bo^*],  
 \label{ham}
 \end{align}
where  $f_\bo=1 + e^{-ik_a}+e^{-i(k_a +k_b)}$, $\hat a=\sqrt{3}\hat x$ and $\hat b= -\sqrt{3}\hat x/2 + 3\hat y/2$ with $k_a=\bo\cdot\hat a$ and $k_b=\bo\cdot\hat b$. Here,  $\boldsymbol{\tau}$ and $\boldsymbol{\sigma}$ are Pauli matrices acting on the layer space and sublattice space respectively. Whereas $I_\tau$ and $I_\sigma$ are $2\times 2$ identity matrix  in each space and $\sigma_\pm= (\sigma_{x} \pm i \sigma_{y})/2$. In order to diagonalize the Hamiltonian, we perform the following canonical ({\it i.e.} commutation relation preserving) transformation of the $\boldsymbol{\sigma}$ and $\boldsymbol{\tau}$ operators
\begin{align}
\sigma_\pm\to \tau_x\sigma_\pm,~ \tau_\pm\to \sigma_x\tau_\pm.
\end{align}
The Hamiltonian Eq.~\ref{ham} now takes the form

 \begin{align}
 \mathcal{H}_\bold k=(E_0 I_\tau - J_L S \tau_x)\otimes I_\sigma -JSI_\tau\otimes[\sigma_+ f_\bo + \sigma_- f_\bo^*].  
 \label{ham1}
 \end{align}
Next, we  diagonalize $(E_0 I_\tau - J_L S \tau_x)$ and the resulting Hamiltonian is given by
 \begin{align}
 \mathcal{H}_\bold k^\alpha=[E_0  +(-1)^{\alpha} |J_L S|] I_\sigma -JS[\sigma_+ f_\bo + \sigma_- f_\bo^*], 
 \label{ham2}
 \end{align}
 where $\alpha=1,2$ for the top and bottom layers respectively. 
The eigenvalues of Eq.~\ref{ham2} yield the magnon bands given in Eq.~\eqref{band}.  The corresponding normalized eigenvectors are tensor product of the two spaces, {\it i.e.} \bea 
\psi_{\bo\pm}^{\alpha}=
\frac{1}{\sqrt 2}{{1}\choose {(-1)^\alpha}}\otimes \frac{1}{\sqrt 2}{{1}\choose {\pm e^{i\phi_\bo}}},
\eea 
 where $\phi_{\bo}=\tan^{-1}\lb{\text{Im} f_\bo}/{\text{Re} f_\bo}\rb$; 
 $\text{Re}$ and $\text{Im}$ denote the real and imaginary parts. 
 
The inversion operation changes the sign of the momentum and interchanges the two sublattices in each layer:

\begin{align}
\mathcal P: \mathcal H_\bo \to I_\tau\otimes \sigma_x \mathcal H_{-\bo} I_\tau\otimes \sigma_x.
\end{align}
It follows that Eq.~\ref{ham} is invariant under inversion symmetry $\mathcal P$. Similar to graphene \cite{fu}, at the time-reversal-invariant momenta (TRIM) points the Hamiltonian has the form
\begin{align}
\mathcal H(\bo=\Gamma_i)= d_0 + d_1(\bo=\Gamma_i)\mathcal P,
\end{align}
where $d_1(\bo)= -JS\text{Re}f_\bo$. Therefore the parity eigenvalues $\xi_n$ for the states at $\bo=\Gamma_i$ are given by the eigenvalues of $\mathcal P$. Hence, $\delta_i=-\text{sgn}\lb d_1(\bo=\Gamma_i)\rb$ and the $\mathbb Z_2$ invariance $\nu$ is given by \cite{fu}
\begin{align}
(-1)^\nu=\prod_{i=1}^4\delta_i.
\end{align}
Indeed,  the DM  spin-orbit interaction has no effects on the values $\delta_i$. It only ensures that a topological gap exists in the system. However, in the absence of any topological gap the $\mathbb Z_2$ invariance $\nu$ defines the topological protection of the 2D nodal-line loops \cite{yang,che}.


\textbf{Acknowledgements}. Research at Perimeter Institute is supported by the Government of Canada through Industry Canada and by the Province of Ontario through the Ministry of Research
and Innovation. 

\textbf{Author contributions}. S. A. Owerre conceived the idea, performed the calculations, discussed the results, and wrote the manuscript.

\textbf{Additional information}

\textbf{Competing financial interests}.  The author declares no competing financial interests.

\end{document}